\documentclass[10pt]{article}
\usepackage{fullpage}
\usepackage{amsmath}
\usepackage{amssymb}
\usepackage{amsfonts}
\usepackage{comment}
\usepackage{color}
\usepackage{cite}
\usepackage{subcaption}
\usepackage{graphicx}
\usepackage{enumitem}

\usepackage{lipsum}

\def\##1{{\bf #1}}
\def\=#1{\underline{\underline{#1}}}

\def\+
#1{\underline{\bf #1}}
\def\*#1{\underline{\underline{\bf #1}}}

\def\r#1{(\ref{#1})}
\def\l#1{\label{#1}}

\def\le{\left(}
\def\ri{\right)}
\def\les{\left[}
\def\ris{\right]}
\def\lec{\left\{}
\def\ric{\right\}}
\def\lek{[{\kern 0.1em}}
\def\rik{{\kern 0.1em}]}

\def\.{\mbox{ \tiny{$^\bullet$} }}

\def\eps{\varepsilon}

\def\ux{\hat{\#u}_{\rm x}}
\def\uy{\hat{\#u}_{\rm y}}
\def\uz{\hat{\#u}_{\rm z}}

\def\Lt{L_{\rm t}}

\def\epst{\eps_{\rm t}}
\def\epsx{\eps_{\rm x}}
\def\epsy{\eps_{\rm y}}
\def\epsz{\eps_{\rm z}}

\def\epsa{\eps_{\mathcal A}}
\def\epsb{\eps_{\mathcal B}}
\def\fa{f_{\mathcal A}}
\def\fb{f_{\mathcal B}}

\newcommand{\norm}[1]{\left\lVert#1\right\rVert}

\begin{document}

\begin{center}

\LARGE{ {\bf Electromagnetic homogenization of  particulate composite materials comprising spheroids and truncated  spheroids with 
orientational
distribution  
}}
\end{center}
\begin{center}
\vspace{10mm} \large
 
  {H\'ector M. Iga-Buitr\'on}\\
{\em School of Mathematics and
   Maxwell Institute for Mathematical Sciences\\
University of Edinburgh, Edinburgh EH9 3FD, UK}
 \vspace{3mm}\\
 {Tom G. Mackay}\footnote{E--mail: T.Mackay@ed.ac.uk.}\\
{\em School of Mathematics and
   Maxwell Institute for Mathematical Sciences\\
University of Edinburgh, Edinburgh EH9 3FD, UK}\\
and\\
 {\em NanoMM~---~Nanoengineered Metamaterials Group\\ Department of Engineering Science and Mechanics\\
The Pennsylvania State University, University Park, PA 16802--6812,
USA}
 \vspace{3mm}\\
 {Akhlesh  Lakhtakia}\\
 {\em NanoMM~---~Nanoengineered Metamaterials Group\\ Department of Engineering Science and Mechanics\\
The Pennsylvania State University, University Park, PA 16802--6812, USA}\\
and\\
{\em School of Mathematics,
University of Edinburgh, Edinburgh EH9 3FD, UK}

\normalsize

\end{center}

\begin{center}
\vspace{5mm} {\bf Abstract}
\end{center}

 Implementations of the Bruggeman and Maxwell Garnett homogenization formalisms were developed to estimate the relative permittivity dyadic of a homogenized composite material (HCM), namely
  $\=\eps^{\rm HCM}$,
  arising from randomly distributed mixtures of electrically-small  particles with spheroidal shapes and truncated spheroidal  shapes.
 The two/three-dimensional (2D/3D) orientational distributions of the component particles were specified by 
  a  Gaussian probability density function.
Numerical investigations were undertaken to explore the relationship between the anisotropy of the HCM and the
 standard deviation of the orientational distribution. 
 For 2D distributions of orientation,  $\=\eps^{\rm HCM}$ is generally biaxial  but it becomes uniaxial when the standard deviation approaches zero or exceeds 3.
For 3D distributions of orientation,  $\=\eps^{\rm HCM}$ is generally uniaxial; however,
  it becomes isotropic when the standard deviation exceeds unity, with  greater degrees of HCM anisotropy arising at smaller values of standard deviation.
The estimates of $\=\eps^{\rm HCM}$ delivered by the Bruggeman formalism and the Maxwell Garnett formalism are in broad agreement, 
over much of the volume-fraction range appropriate to the Maxwell Garnett formalism,
but the degree of HCM anisotropy predicted by the Maxwell Garnett formalism is generally a little higher
than that predicted by the Bruggeman formalism, especially at low values of standard deviation.

 \vspace{5mm}
 
\vspace{5mm}

\section{Introduction}

Composite materials composed of random distributions of electrically-small particles provide the setting for this paper.  Electromagnetically, 
the particulate composite material 
may be regarded as a homogeneous material,
assuming that
the component particles are much smaller than the wavelengths involved \cite{SelPap, Choy,Qin_JAP}.
The process of representing a composite material 
as a homogeneous material in the
 long-wavelength
regime
 is called \emph{homogenization} and the 
 homogeneous material itself is called the \emph{homogenized composite material} (HCM).
Through the process of homogenization, constitutive parameters
may be extended \cite{Neelakanta}; in certain instances, even  entirely new  constitutive parameters
may emerge \cite{Meet}.
For example, an HCM arising from component particles made of  isotropic dielectric materials may itself be an anisotropic 
dielectric material
if the component particles are shaped and oriented appropriately. Accordingly, HCMs can
play important roles in applications. Notably, 
nanocomposite materials have been harnessed for many
 recent and ongoing advances in optical applications \cite{Werd,Pinar}.

Numerous formalisms have been developed to estimate the constitutive parameters of HCMs.
Two of the most widely used 
are the Bruggeman formalism \cite{Br,Ward} and the Maxwell Garnett formalism \cite{MG1904,Markel}. 
Implementations of both of these formalisms have been established for the most general linear HCMs \cite{WLM_MOTL}. 
 The Maxwell Garnett formalism has the advantage of computational simplicity, but it is limited to small volume fractions of component particles. On the other hand, the Bruggeman formalism is not restricted to small volume fractions of component particles, but it is somewhat more computationally expensive \cite{MAEH}.
 The component particle shapes accommodated by homogenization formalisms are usually spherical, spheroidal, or elliptical, but this palette of component particle shapes has recently been extended to include truncated spheroids \cite{ML_WRCM,TruncatedSpheroids} and superspheroids \cite{Superspheroids}. 

In 
most implementations of
homogenization formalisms,
the particles of each component material   are  taken to be identically oriented. In some instances, such an alignment of component particles could be achieved through the application of an external unidirectional electric field \cite{Park} or magnetic field \cite{Erb}; in other instances, alignment can be flow-induced \cite{Kunji}.
 In practice, the alignment of component particles may only be partially achieved. Indeed, for certain applications, it may be desirable to tune the anisotropy exhibited by the HCM through varying the degree of alignment of the component particles.
Ranges of particle orientation have been highlighted in  studies based on 
 homogenization of
  dilute mixtures of particles with simple shapes \cite{MG_PIER,Kfoury}.
 The prospect of composite materials based on complex-shaped component particles that are non-uniformly oriented offers greater opportunities for technologists~--~and also greater challenges for theorists.

 In the following, the Bruggeman formalism  and the Maxwell Garnett formalism  are developed for an HCM based on component particles shaped as truncated spheroids
 \cite{ML_WRCM,TruncatedSpheroids}, the derived expressions being straightforwardly extensible to untruncated spheroids.
 The component particles are oriented according to a  Gaussian probability density function.
 Orientation distributional distributions in a plane (i.e., 2D) and in a volume (i.e., 3D) are both considered.
  The shapes and 
  orientations of the component particles render the HCM 
  anisotropic. The HCM anisotropy may be either uniaxial or biaxial, depending on the orientational distribution
  of the component particles. The influence of the distribution of component particle orientations upon the anisotropy of the HCM is explored in numerical studies. Additional numerical results, as well as the 
  the 
  MATLAB codes used to generate the numerical results, are provided in the {\bf Supplementary Material}.

As regards notation, 
vectors are represented in  boldface and 3$\times$3 dyadics \cite{Chen} are denoted by double underlining.
The  unit vectors
$\ux$, $\uy$, and $\uz$ are
 parallel to the  coordinate axes of the Cartesian coordinate system $(x,y,z)$.
   The identity dyadic is written as $\=I=   \ux\ux+ \uy\uy+ \uz\uz $ and the null dyadic is
   written as
    $\=0$.
  Also,  
  \begin{equation}
  {\rm erf}(z) = 
  \frac{2}{\sqrt{\pi}} \int^z_0 \exp \le - \tau^2 \ri \, d \tau
  \end{equation} is the Gaussian error function,
   angular frequency  is  denoted by $\omega$, and $i = \sqrt{-1}$. 
  
\section{Preliminaries: component materials and particles} \l{Prelim_Sec}

Consider a composite material comprising a  mixture of particles made from two different 
materials, one  labelled $\mathcal{A}$ and the other labelled $\mathcal{B}$. 
Material $\mathcal{A}$ is a homogeneous isotropic dielectric material of relative permittivity $\epsa$ and material $\mathcal{B}$ is a  homogeneous isotropic dielectric material of relative  permittivity $\epsb$. 
The volume fraction of material  $\mathcal{A}$   is denoted by $\fa\in[0,1]$ while that of material $\mathcal{B}$ by $\fb = 1-\fa$.
Particles of materials $\mathcal{A}$ and $\mathcal{B}$  are randomly distributed in space with all particles of $\mathcal{A}$   having the same shape and, likewise, all particles of   $\mathcal{B}$   having the same shape, 
but the shape of the particles of material  $\mathcal{A}$ being different from the shape of the particles of $\mathcal{B}$.

\begin{figure}[!htb]
\centering
\includegraphics[width=12cm]{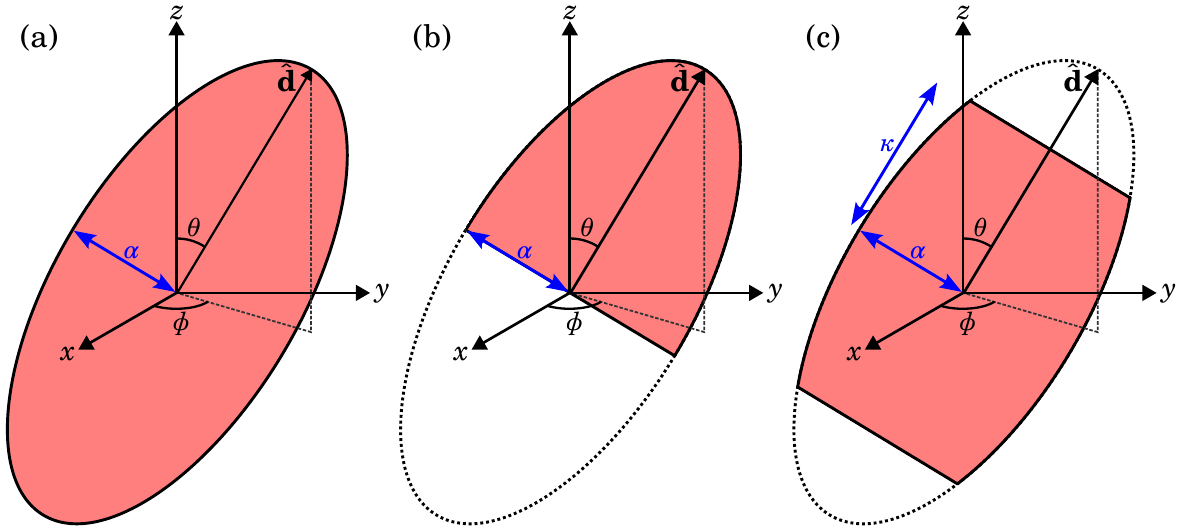} 
  \vspace{0mm}  \hfill
 \caption{\label{Fig1} Particles whose
 shapes  are based on a spheroid
   with equatorial radius $\alpha$ and unit polar radius: (a) spheroid, (b) hemispheroid, and
    (c) doubly-truncated spheroid. The unit vector aligned with the rotational axis of symmetry is $\hat{\#d}
    = \ux \sin \theta \, \cos \phi+\uy \sin \theta \, \sin \phi+\uz\cos \theta
    $.}
\end{figure}

The following three different shapes are considered separately for the particles of $\mathcal{A}$:
\begin{enumerate}
\item spheroidal,
\item hemispheroidal, and
\item doubly-truncated spheroidal.
\end{enumerate}
All three are illustrated   in Fig.~\ref{Fig1}. Each shape is based on a spheroid
with an equatorial radius $\alpha$ and a unit polar radius,  with its axis of rotational symmetry 
  aligned with the unit vector $\hat{\#d}
    =\ux \sin \theta \, \cos \phi+\uy \sin \theta \, \sin \phi+\uz\cos \theta $.
   The
  hemispheroidal shape arises from the spheroidal shape through truncation by the plane perpendicular to the polar axis. The
  doubly-truncated spheroidal shape arises from the spheroidal shape through truncation by two planes,
  both perpendicular to the polar axis, equidistant from the truncation plane used for the hemispheroidal shape.
The distance between the two truncation planes for the doubly-truncated spheroidal shape is $2 \kappa$.
The orientation of each   particle is specified by the polar angle $\theta$ relative to the $z$ axis
and the azimuthal angle $\phi$ relative to the $x$ axis in the $xy$ plane.
The  orientation specified by $(\theta, \phi)$ is indistinguishable from the 
 orientation specified by $(\pi - \theta, \phi + \pi)$ for both spheroidal particles and doubly-truncated spheroidal  
 particles, but that is untrue for
hemispheroidal particles.
The orientational distribution of the particles of
material $\mathcal{A}$   is governed by a probability density function (PDF) specified in the later sections.

\begin{figure}[!htb]
    \centering
    \includegraphics[width=8cm]{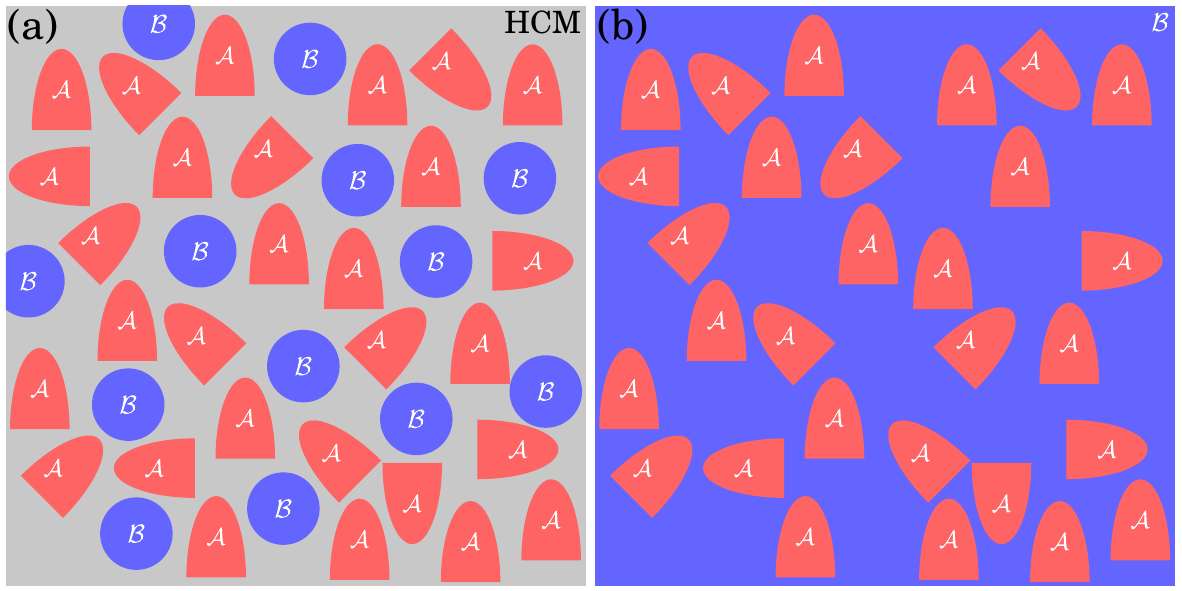}
    \caption{Schematic representation of the particulate composite material for the (a) Bruggeman formalism and (b) Maxwell Garnett formalism, for the case in which the particles of material $\mathcal{A}$  
 are  hemispheroidal.}
    \label{Fig2}
\end{figure}

Provided that all particles are electrically small, i.e., at least 10 times smaller than the wavelengths involved \cite{van}, the composite material may be regarded as an HCM whose constitutive parameters may be estimated by means of a homogenization formalism. 
The Maxwell Garnett formalism
is agnostic about the shapes of the particles of material $\mathcal{B}$. For simplicity, the particles of material $\mathcal{B}$   are taken to be  spheres in the case of the Bruggeman formalism.   Therefore no orientation is assigned to these particles in the remainder
of this paper.   
Schematic representations of the component particles in the particulate
composite material are provided in Fig.~\ref{Fig2} for both formalisms, for the case in which the particles of material $\mathcal{A}$  
 are  hemispheroidal.

\section{2D orientational distribution of particles of material $\mathcal{A}$ }

Let us begin with the case in which the polar axes for all particles of material $\mathcal{A}$   lie wholly in the same plane; i.e.,
we have a 2D orientational distribution of these particles.
Without loss of generality, this plane is taken as the $xz$ plane so that   $\phi \equiv 0$.
In this setting, we consider only spheroidal particles of material $\mathcal{A}$.

\subsection{Smooth distribution of   orientations in the $xz$ plane} \l{smooth_sec_2d}

Let $\mathcal{S}_{\mathcal{A}}$ denote the set of all particles of material
$\mathcal{A}$.
We consider $\mathcal{S}_{\mathcal{A}}$ to be  partitioned as
\begin{equation} \l{A_part}
    \mathcal{S}_{\mathcal{A}} = \mathcal{S}_{\mathcal{A}}^{[\theta_0,\theta_1]} \cup \mathcal{S}_{\mathcal{A}}^{[\theta_1,\theta_2]} \cup \ldots \cup \mathcal{S}_{\mathcal{A}}^{[\theta_{N-1},\theta_N]},
\end{equation}
where $\mathcal{S}_{\mathcal{A}}^{[\theta_{k},\theta_{k+1}]}$ represents the subset of 
those particles
 whose polar angles lie in the range $\theta_k < \theta < \theta_{k+1}$, $k\in\lec0, 1, ... , N-1\ric$, with $\theta_0=-\pi/2$ and $\theta_N=\pi/2$. 
 The number $N$ of subsets is at least unity.
The volume fraction occupied by the particles  belonging to the subset $\mathcal{S}_{\mathcal{A}}^{[\theta_{k},\theta_{k+1}]}$ is $\fa^{[\theta_k, \theta_{k+1}]}$;
hence,
\begin{equation}
    \fa = \fa^{[\theta_{0},\theta_{1}]} + \fa^{[\theta_{1},\theta_{2}]} + \ldots + \fa^{[\theta_{N-1},\theta_{N}]}.
\end{equation}
The probability that the polar orientational angle   lies in the range $\theta_k < \theta < \theta_{k+1}$ is
\begin{equation}
    \mathrm{P}[\theta_k < \theta < \theta_{k+1}]= \displaystyle
    \frac{ \fa^{[\theta_k,\theta_{k+1}]}
 }{\fa}   
    =
    \int_{\theta_k}^{\theta_{k+1}} g_2(\theta) \,  \mathrm{d}\theta,
\end{equation}
wherein the PDF $g_2(\theta)$ is normalized as follows:
\begin{equation} \l{g_norm}
    \int_{-\pi/2}^{\pi/2} g_2(\theta) \, \mathrm{d}\theta=1.
\end{equation}

The truncated Gaussian PDF
\begin{equation}\label{eq_2Dpdf}
g_2(\theta)=\frac{1}{\eta_2} \exp \left( -\frac{\theta^2}{2 \sigma^2} \right)
\end{equation}
is adopted here, with $\sigma$ being the standard deviation and  the normalization constant 
\begin{equation}
    \eta_2 = \sqrt{2\pi} \ \sigma \ \mathrm{erf}\left( \frac{\pi}{2 \sqrt{2} \sigma} \right) ,
\end{equation}
in conformity with the  constraint \r{g_norm}.
Thus,
the most probable polar orientation angle for these particles is $\theta = 0$, whereas
the least  probable polar orientation angle for these particles is $\theta = \pi/2$ (or equivalently  
$\theta = -\pi/2$). 
All  particles get aligned with the $z$ axis  as $\sigma \to 0$; in contrast,
all polar orientation angles become equally probable   as $\sigma \to \infty$.

\subsection{Piecewise-uniform distribution of   orientations in the $xz$ plane}\l{PU_2d}

For the implementation of homogenization formalisms, 
the smooth distribution of 
particle orientation angles described in \S\ref{smooth_sec_2d}
is approximated by
 a piecewise-uniform distribution. 
For this purpose,  $N$ is taken to be sufficiently large that the partition \r{A_part} may be
 replaced by the partition
\begin{equation}
    \mathcal{S}_{\mathcal{A}} = \mathcal{S}_{\mathcal{A}}^{\left[ \bar{\theta}_0 \right]} \cup \mathcal{S}_{\mathcal{A}}^{\left[ \bar{\theta}_1 \right]} \cup \ldots \cup \mathcal{S}_{\mathcal{A}}^{\left[ \bar{\theta}_{N-1} \right]},
\end{equation}
where 
$\mathcal{S}_{\mathcal{A}}^{\left[ \bar{\theta}_{k} \right]}$ represents the subset of   
particles 
 whose polar orientation angle is $\bar{\theta}_{k} = (\theta_k+\theta_{k+1})/2 $, $k\in\lec0, 1, ... , N-1\ric$.
The volume fraction of 
  particles
 belonging to the subset $\mathcal{S}_{\mathcal{A}}^{\left[ \bar{\theta}_{k} \right]}$ is denoted as $f_{\mathcal{A}_k}$. 
For this piecewise-uniform distribution, the probability that a
  particle has a polar orientation angle $\bar{\theta}_k$ is given as
\begin{equation}
    \mathrm{P}\left[ \theta = \bar{\theta}_k \right] =
    \frac{  f_{\mathcal{A}_k}}{\fa} =
     g_2 \left( \bar{\theta}_k \right) \, \Delta\theta_k,
\end{equation}
with $\Delta\theta_k = \theta_{k+1} - \theta_{k}$. 
And  $N$ is taken to be sufficiently large that the constraint
\begin{equation}
    \displaystyle\sum_{k=0}^{N-1} g_2 \left( \bar{\theta}_k \right) \, \Delta\theta_k = 1 
\end{equation}
holds.

\subsection{Sampling of PDF: 2D   orientational distribution} \l{Sample2D}

 Now we address the question: At which values  $\bar{\theta}_k \in \le -\pi/2, \pi/2 \ri$ should the PDF $g_2 (\theta)$ be sampled so that the piecewise-uniform distribution of  orientations described in \S\ref{PU_2d}  adequately represents the smooth distribution of
 orientations described in \S\ref{smooth_sec_2d}? In order to take into account   the  Gaussian nature of 
 $g_2(\theta)$, the sampling density, as gauged by
  $\le \Delta\theta_k \ri^{-1}$,  is chosen to be  proportional to the value of $g_2(\bar{\theta}_k)$. In particular, 
   the greatest  sampling density  arises at that value of $\bar{\theta}_k$  where  $g_2 (\bar{\theta}_k)$  has its maximum value.
This outcome is  achieved by employing the inverse transform sampling method \cite{ITSM_Hormann,ITSM_Little,ITSM_Umeda},
which exploits the  relationship between $g_2 (\theta)$ 
defined in Eq.~\r{eq_2Dpdf}
 and its cumulative distribution function  (CDF) 
\begin{equation}\label{eq_2Dcdf} 
G_2(\theta) = \displaystyle\int_{-\pi/2}^{\theta} g_2(\theta') \, \mathrm{d}\theta'=\frac{1}
{2} \left[
1 + \frac{\mathrm{erf}\left( \frac{\theta}{\sqrt{2}\sigma} \right)}{\mathrm{erf}\left( \frac{\pi}{2\sqrt{2}\sigma} \right)} \right].
\end{equation}
Thus,  $G_2(\theta)$ is a monotonically increasing function that maps the domain of  $g_2 (\theta)$, i.e., $[-\pi/2, \pi/2]$, onto the interval $[0,1]$,  such that $G_2(-\pi/2)=0$ and $G_2(\pi/2)=1$. 

\begin{figure}[!htb]
\centering
\includegraphics[width=16cm]{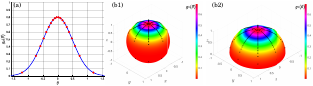} 
  \vspace{0mm}  \hfill
 \caption{\label{Fig3} 
(a)  The PDF $g_2 (\theta)$ plotted against $\theta \in \le -\pi/2, \pi/2 \ri$
 with $\sigma=0.5$. 
 The sampled values $\bar{\theta}_k$, $k\in\lec0, 1, ... , N-1\ric$,   of $\theta$ are identified as red dots for $N=20$. (b1,b2) The  PDF $g_3 (\theta)$  plotted on the surface of the unit sphere 
 for  (b1) $\theta_N = \pi$  and (b2) $\theta_N  = \pi/2$,  with $\sigma=0.5$. The sampled values   $\bar{\theta}_k$, $k\in\lec0, 1, ... , N-1\ric$, 
 of $\theta$ and the sampled values
 $\bar{\phi}_\ell$, $\ell\in\lec0, 1, ... , M-1\ric$,
of $\phi$ are identified as black dots for $N=20$ and $M=8$. }
\end{figure}

 The sampled values $\bar{\theta}_k$ for {$k\in\lec0, 1, \ldots,N-1\ric$} are found as follows. First,
the set of $N$ numbers $\lec q_0, q_1, \ldots, q_{N-1} \ric$ is generated that  uniformly spans
 the interval $(0,1)$. The values of $q_k$, {$k\in\lec0, 1, \ldots,N-1\ric$}, demarcate uniformly-spaced levels of cumulative probability. Then, for each  value of $q_k$, the corresponding value  $\bar{\theta}_k$ is provided via
\begin{equation} \l{G2}
G_2(\bar{\theta}_k)=q_k, \qquad k\in\lec0, 1, ... , N-1\ric.
\end{equation}
In order to extract $\bar{\theta}_k$ from Eq.~\r{G2}, it is necessary to compute the inverse of  $G_2(\bar{\theta}_k)$, which  can be achieved using 
numerical methods such as interpolation \cite{Steffensen}.
This process concentrates  sampling in $\theta$-regions where $g_2 (\theta)$  has higher values, reflecting the  distribution of    orientations, as illustrated in Fig.~\ref{Fig3}(a).

Sampling is sensitive to the value of $\sigma$. For values of $\sigma$ close to zero, there is a high density of sampling  in the neighborhood of the maximum of $g_2 (\theta)$, reflecting the sharp peak of the  Gaussian distribution. As  $\sigma$ increases, the truncated Gaussian distribution becomes broader  and the density of sampling within the domain $ \le-\pi/2, \pi/2\ri$  becomes more uniform.

\subsection{Homogenization formalisms: 2D   orientational distribution} \l{homog_2d_sec}

Owing to the distribution of orientations of the particles of $\mathcal{A}$   parallel to the $xz$ plane, the
relative permittivity dyadic of the HCM, namely $\=\eps^{\rm HCM}$, generally has the biaxial
form
\begin{equation} \l{eps-biaxial}
\=\eps^{\rm HCM} = \epsx^{\rm HCM} \, \ux \, \ux + \epsy^{\rm HCM} \, \uy \, \uy + \epsz^{\rm HCM} \, \uz \, \uz,
\end{equation}
with
the relative permittivity parameters
 $\epsx^{\rm HCM}\in\mathbb{C}$, $\epsy^{\rm HCM}\in\mathbb{C}$, and $\epsz^{\rm HCM} \in\mathbb{C}$.
The  Bruggeman estimate of $\=\eps^{\rm HCM}$ is written as $\=\eps^{\rm Br}= \epsx^{\rm Br} \, \ux \, \ux + \epsy^{\rm Br} \, \uy \, \uy + \epsz^{\rm Br} \, \uz \, \uz$ and the Maxwell Garnett estimate of 
$\=\eps^{\rm HCM}$ is written
as $\=\eps^{\rm MG}= \epsx^{\rm MG} \, \ux \, \ux + \epsy^{\rm MG} \, \uy \, \uy + \epsz^{\rm MG} \, \uz \, \uz$. 

\subsubsection{Bruggeman homogenization formalism:  2D orientational distribution } \l{Br_sec_2d}

The  Bruggeman estimate  $\=\eps^{\rm Br}$ is provided implicitly by the nonlinear dyadic equation
\cite{MAEH}
\begin{equation} \label{Br_2d}
\le
\sum^{N-1}_{k=0}  f_{\mathcal{A}_k} \, \=a^{\mathcal{A}_k/ \rm Br} \ri + \fb \, \=a^{\rm \mathcal{B}/Br}=\=0.
\end{equation}
Herein the polarizability density dyadics
\begin{equation}  \label{a_2d}
 \=a^{\rm j/Br} = \le \eps_\mathcal{A} \=I - \=\eps^{\rm Br} \ri \.   \=P^{\rm j/Br} , \qquad {\rm j} \in \lec
 \mathcal{A}_k, \mathcal{B} \ric,
 \end{equation}
are defined using the dyadics
 \begin{equation} \l{p_2d}
 \left.
 \begin{array}{l}
 \=P^{ \mathcal{A}_k/ \rm Br} =  \les \=I + i \omega \=D^{\mathcal{A}_k/ \rm  Br} \. \le \eps_\mathcal{A} \=I - \=\eps^{\rm Br} \ri \ris^{-1} \vspace{4pt}\\
 \=P^{\rm\mathcal{B}/Br} =  \les \=I + i \omega \=D^{\rm \mathcal{B}/Br} \. \le \eps_\mathcal{B} \=I - \=\eps^{\rm Br} \ri \ris^{-1}
 \end{array}
 \right\},
 \end{equation}
which contain the depolarization dyadic $\=D^{\mathcal{A}_k/ \rm  Br}$ relevant
to a particle of material $\mathcal{A}$   with polar orientation angle
$\bar{\theta}_k$
 immersed in the  HCM, 
 and the depolarization dyadic $\=D^{\rm \mathcal{B}/Br}$
for a particle of material $\mathcal{B}$   
 immersed in the  HCM. Expressions for these depolarization dyadics are provided in the Appendix.
 
 A Jacobi iterative scheme \cite{Jacobi} is employed to numerically extract  $\=\eps^{\rm Br}$ from Eq.~\r{Br_2d}, as follows.
 The $\le n+1 \ri^{\rm th}$ iterate of  $\=\eps^{\rm Br}$ is delivered in terms of its
  $n^{\rm th}$ iterate as
\begin{equation} \l{Jacob_2d}
    \=\eps^{\mathrm{Br}}[n+1] =  \les  \eps_{\mathcal{A}}  \left(\displaystyle\sum_{k=0}^{N-1}  f_{\mathcal{A}_k} \, \={P}^{\mathcal{A}_k/ \rm  Br}[n] \right) + \fb \, \eps_{\mathcal{B}} \, \={P}^{\rm \mathcal{B}/Br}[n]  
   \ris \. \les \left( \displaystyle\sum_{k=0}^{N-1} f_{\mathcal{A}_k} \, \={P}^{\mathcal{A}_k/ \rm Br}[n] 
   \right)
   +  \fb \, \={P}^{\rm \mathcal{B}/Br}[n] \ris^{-1},
\end{equation}
wherein
 the dyadics
 \begin{equation} \l{PAPB}
 \left.
 \begin{array}{l}
 \=P^{\mathcal{A}_k/ \rm  Br}[n] =  \les \=I + i \omega \=D^{{\mathcal{A}_k/ \rm Br}}[n] \. \le \eps_\mathcal{A} \=I - \=\eps^{\rm Br}[n] \ri \ris^{-1} \vspace{4pt}\\
 \=P^{\rm\mathcal{B}/Br}[n] =  \les \=I + i \omega \=D^{{\rm \mathcal{B}/Br}}[n] \. \le \eps_\mathcal{B} \=I - \=\eps^{\rm Br}[n] \ri \ris^{-1}
 \end{array}
 \right\},
 \end{equation}
with $ \=D^{{\mathcal{A}_k/ \rm Br}}[n]$ and  $\=D^{{\rm \mathcal{B}/Br}}[n]$ being defined as
$ \=D^{{\mathcal{A}_k/ \rm Br}}$ and $\=D^{{\rm \mathcal{B}/Br}}$, respectively, but for particles immersed in the medium specified by the relative permittivity dyadic $\=\eps^{\rm Br}[n] $.
In practice, $\=\eps^{\mathrm{Br}}[0] = \fa\, \=\eps_{\mathcal{A}} + \fb \,\=\eps_{\mathcal{B}}$ is found
to be a suitable 
 initial dyadic. Typically, the Jacobi scheme    converged adequately within 12 iterations for all results presented in \S\ref{Results2D}.

\subsubsection{Maxwell Garnett homogenization formalism: 2D orientational distribution } \l{MG_sec_2d}

The  Maxwell Garnett   estimate of $\=\eps^{\mathrm{HCM}}$ is delivered explicitly as
\cite{MAEH}
\begin{equation}
     \=\eps^{\mathrm{MG}}= \eps_{\mathcal{B}}\, \={I} + \displaystyle\sum_{k=0}^{N-1} \left[ f_{\mathcal{A}_k} \, \={a}^{\mathcal{A}_k / \mathcal{B}}  \. \left(\={I} -   \frac{f_{\mathcal{A}_k}}{3 \eps_{\mathcal{B}}} \, \={a}^{\mathcal{A}_k/\mathcal{B}} \right)^{-1} \right],
\end{equation}
wherein 
the polarizability density dyadic
\begin{equation} \l{a_MG_2d}
  \={a}^{\mathcal{A}_k / \mathcal{B}} = \le \eps_\mathcal{A}- \eps_\mathcal{B} \ri  
   \les \=I + i \omega  \le \eps_\mathcal{A}  - \eps_{\mathcal{B}} \ri \=D^{\mathcal{A}_k/\mathcal{B}}  \ris^{-1}
 \end{equation}
 contains the depolarization dyadic $\=D^{ \mathcal{A}_k/\mathcal{B}}$
relevant for a particle of material   $\mathcal{A}$ particle with polar orientation angle
$\bar{\theta}_k$
 immersed in material $\mathcal{B}$. An expression for this depolarization dyadic is provided in the Appendix.

\subsection{Numerical results:  2D orientational distribution} \l{Results2D}

\begin{figure}[!htb]
\centering
\includegraphics[width=14cm]{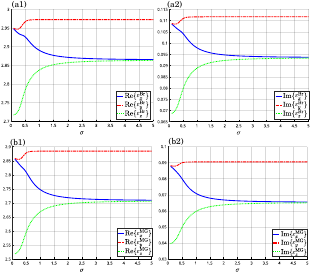} 
  \vspace{0mm}  \hfill
\caption{\label{Fig4} Real and imaginary components of (a) $\underline{\underline{\eps}}^{\rm Br}$ and (b) $\underline{\underline{\eps}}^{\rm MG}$ versus 
 $\sigma$ for spheroidal particles of material $\mathcal{A}$ oriented in the $xz$ plane, when $\fa = 0.3$ and 
$\alpha=3$.}
\end{figure}

The estimates  $\=\eps^{\mathrm{Br}}$  and $ \=\eps^{\mathrm{MG}}$ provided in 
\S\ref{Br_sec_2d} and \S\ref{MG_sec_2d} are now numerically investigated for
$\epsa = 6.0 + 0.5 i$ and $\epsb = 2.0 + 0.01 i$, which are representative of many commonly encountered dissipative dielectric materials \cite{Wakaki}.
 The  orientations of the particles of material  $\mathcal{A}$  
are distributed in the $xz$ plane, per the  Gaussian PDF in Eq.~\r{eq_2Dpdf}.

We focus on the
relationship between
  the  components of the relative permittivity dyadics $\=\eps^{\rm Br}$ and $\=\eps^{\rm MG}$, and  the standard deviation $\sigma$ that characterizes the orientational distribution of the particles of material $\mathcal{A}$.
  Numerical results are presented
  for  $\fa = 0.3$  and   $\alpha = 3$.
Note that  the Maxwell Garnett formalism is  appropriate only for dilute composite materials
with $\fa \lessapprox 0.3$ \cite{SelPap,Niklasson,Markel}, but there is no such limitation on the
Bruggeman formalism.

The number $N$ of sampled values
 $\bar{\theta}_k$, {$k\in\lec0, 1, \ldots,N-1\ric$}, was determined by 
 numerical
 experimentation as follows. Repeated computations were undertaken with $N$ being gradually increased until the  relative infinity norms of 
 $\=\eps^{\rm Br}$ and $\=\eps^{\rm MG}$ were acceptably small. Specifically, $N$ was set by requiring that the tolerances $\delta^{\rm j} < 0.0001$ were met, where
 \begin{equation}\l{eq_Error}
 \delta^{\rm j} =  \frac{\norm{  \=\eps^{\rm j}_{[v+1]} - 
  \=\eps^{\rm j}_{[v]}}_\infty}{\norm{\=\eps^{\rm j}_{[v+1]} }_\infty}  , \qquad  {\rm j \in\left\{ Br, MG\right\}}, 
 \end{equation}
 with $\=\eps^{\rm j}_{[v]}$
 being the $v^{\rm th}$ computation  of $\=\eps^{\rm j}$ and $\norm{\cdot}_\infty$ being the infinity norm. Typically, $N=60$ was found to be adequate for small values of $\sigma$, and  smaller values of $N$ could be used for larger values of $\sigma$.

The real and imaginary parts of the components of $\=\eps^{\rm Br}$ and $\=\eps^{\rm MG}$ are plotted as functions of the standard deviation $\sigma \in [0.1,5]$ in Fig.~\ref{Fig4}.
As discussed in \S\ref{homog_2d_sec}, due to the 2D distribution of orientations of the particles of material $\mathcal{A}$, the relative permittivity dyadic of the HCM generally takes the biaxial form given by Eq.~\r{eps-biaxial}  for both Bruggeman and Maxwell Garnett formalisms. 
In the limit $\sigma \to 0$,  the rotational symmetry axes of all  particles align along the $z$ axis and
the HCM becomes uniaxial
with
$\eps_{\rm x}^{\rm Br} = \eps_{\rm y}^{\rm Br}$ and $\eps_{\rm x}^{\rm MG} = \eps_{\rm y}^{\rm MG}$. 
As $\sigma$ increases, $\eps_{\rm y}^{\rm Br}$ and 
$\eps_{\rm y}^{\rm MG}$
 remain approximately constant,  $\eps_{\rm x}^{\rm Br}$ 
 and $\eps_{\rm x}^{\rm MG}$
 decrease, and $\eps_{\rm z}^{\rm Br}$ 
 and $\eps_{\rm z}^{\rm MG}$
 increase. 
 As $\sigma$ increases beyond $3$,
 the orientations of the   particles of material $\mathcal{A}$ become progressively closer to being uniformly distributed 
 in the $xz$ plane;
eventually,  the HCM becomes uniaxial with
 $\eps_{\rm x}^{\rm Br} = \eps_{\rm z}^{\rm Br}$, and  $\eps_{\rm x}^{\rm MG} = \eps_{\rm z}^{\rm MG}$.

The plots of the components of $\=\eps^{\rm Br}$ and $\=\eps^{\rm MG}$ in Fig.~\ref{Fig4} are
 qualitatively similar. However, the real parts of the
 components  of $\=\eps^{\rm MG}$ are consistently smaller than those of $\=\eps^{\rm Br}$, and likewise
 the imaginary parts of the
 components  of $\=\eps^{\rm MG}$ are consistently smaller than those of $\=\eps^{\rm Br}$.
This discrepancy between the Bruggeman estimate and the Maxwell Garnett estimate is most conspicuous when the volume fraction $\fa$ is in the neighborhood of the Maxwell Garnett limit $\fa \approx 0.3$ \cite{TruncatedSpheroids,Superspheroids}. For much smaller values of $\fa$, numerical studies (not presented here) reveal close quantitative and qualitative agreement between the estimates of the two formalisms.

 \section{3D orientational distribution of particles of material $\mathcal{A}$}

Next we consider the more general case in which the
polar axes of the particles of material  
  $\mathcal{A}$   are distributed throughout three-dimensional space.
 In this setting, particles of material $\mathcal{A}$ shaped as spheroids, hemispheroids, and doubly-truncated spheroids
 are considered separately,  per Fig.~\ref{Fig1}.

\subsection{Smooth distribution of   orientations in three dimensions} \l{smooth_sec_3d}

As in \S\ref{smooth_sec_2d},
 $\mathcal{S}_{\mathcal{A}}$ denotes the set of all 
{particles of material $\mathcal{A}$}.
We consider $\mathcal{S}_{\mathcal{A}}$ as being  partitioned as
\begin{equation} \l{A_part_3d}
    \mathcal{S}_{\mathcal{A}} = 
    \bigcup^{M-1}_{\ell =0}   \bigcup^{N-1}_{k=0}    \mathcal{S}_{\mathcal{A}}^{[\theta_k,\theta_{k+1}; \phi_\ell, \phi_{\ell+1}]} ,
\end{equation}
where  $\mathcal{S}_{\mathcal{A}}^{[\theta_k,\theta_{k+1}; \phi_\ell, \phi_{\ell+1}]}$ represents the subset of 
{particles of material $\mathcal{A}$}
 whose polar angles lie in the range $\theta_k < \theta < \theta_{k+1}$, $k\in\lec0, 1, \ldots,N-1\ric$,
  and whose azimuthal angles lie in the range $\phi_\ell < \phi < \phi_{\ell+1}$, $\ell\in\lec0, 1, \ldots,M-1\ric$,
 with $\theta_0=0$, $\phi_0 =0$, and  $\phi_M=2 \pi$. 
 The upper bound on the $\theta$-range depends upon the symmetry of the particles of material $\mathcal{A}$  being considered. 
We set $\theta_N  = \pi/2$ in the case of spheroidal and doubly-truncated spheriodal {particles of material $\mathcal{A}$} (which are unchanged under the mapping $\left\{\theta \mapsto \pi - \theta,\,\phi \mapsto \phi + \pi\right\}$). In contrast,
when the {particles of material $\mathcal{A}$ are hemispheroidal} 
(which are changed under the mapping $\left\{\theta \mapsto \pi - \theta,\,\phi \mapsto \phi + \pi\right\}$), we generally set $\theta_N  = \pi$ (but  the effect of setting $\theta_N = \pi/2$   is explored in Fig.~\ref{Fig9}).
 
The volume fraction occupied by {particles of material $\mathcal{A}$}  belonging to the subset $\mathcal{S}_{\mathcal{A}}^{[\theta_{k},\theta_{k+1}; \phi_\ell, \phi_{\ell+1}]}$ is $\fa^{[\theta_k, \theta_{k+1}; \phi_\ell, \phi_{\ell+1}]}$, with
\begin{equation}
    \fa = 
       \sum^{M-1}_{\ell =0}   \sum^{N-1}_{k=0}
  \fa^{[\theta_k, \theta_{k+1}; \phi_\ell, \phi_{\ell+1}]}.
\end{equation}
The probability that the polar orientational angle for a particle of material  $\mathcal{A}$  lies in the range $\theta_k < \theta < \theta_{k+1}$ 
and the azimuthal orientational angle lies in the range $\phi_\ell < \phi < \phi_{\ell+1}$
is
\begin{equation}
    \mathrm{P}[\theta_k < \theta < \theta_{k+1}, \phi_\ell < \phi < \phi_{\ell+1}]= \displaystyle
    \frac{ \fa^{[\theta_k,\theta_{k+1}; \phi_\ell < \phi < \phi_{\ell+1}]}
 }{\fa}   
    =
    \int_{\phi_\ell}^{\phi_{\ell+1}} \int_{\theta_k}^{\theta_{k+1}} g_3(\theta, \phi) \, \sin \theta \, \mathrm{d}\theta \, d \phi,
\end{equation}
wherein the PDF $g_3(\theta, \phi)$ satisfies the constraint
\begin{equation} \l{g_norm_3d}
    \int_{\phi = 0}^{2 \pi} \int_{\theta =0 }^{\theta_N } g_3(\theta, \phi) \,  \sin \theta \, \mathrm{d}\theta \, d \phi=1.
\end{equation}

We focus on orientational distributions that are invariant under rotation about the $z$ axis.
Thus, 
 all azimuthal orientation angles $\phi$ are equally probable for 
{particles of material $\mathcal{A}$}. Accordingly, henceforth we write  $g_3(\theta)$ in lieu of  $g_3(\theta, \phi)$.
The truncated Gaussian PDF
\begin{equation}\label{eq_3Dpdf}
g_3(\theta)=\frac{1}{\eta_{3}} \exp \left( -\frac{\theta^2}{2 \sigma^2} \right)
\end{equation}
is adopted here, with   the normalization constant 
\begin{equation}
    \eta_3  =
    \displaystyle{ \frac{\pi^{3/2} \, \sigma \, \exp\left( -\frac{\sigma^2}{2}\right) }{\sqrt{2}} \,  \left[ - 2i \, \mathrm{erf}\left( \frac{i \, \sigma}{\sqrt{2}} \right) + i \, \mathrm{erf}\left( \frac{i \, \sigma^2 + \theta_N }{\sigma \sqrt{2}} \right) + i \, \mathrm{erf}\left( \frac{i \, \sigma^2 - \theta_N  }{\sigma \sqrt{2}} \right) \right] },
\end{equation}
in conformity with the  constraint \r{g_norm_3d}.
Thus, much like the 2D orientational distribution  described in  \S\ref{smooth_sec_2d},
the most probable polar orientation angle for {particles of material $\mathcal{A}$} is $\theta = 0$ while
the least  probable polar orientation angle for {particles of material $\mathcal{A}$} is $\theta = \theta_N $. In the limit as $\sigma \to 0$, all {particles of material $\mathcal{A}$} are aligned parallel to the $z$ axis;
 in the limit as $\sigma \to \infty$, all polar orientation angles are equally probable.

\subsection{Piecewise-uniform distribution of   orientations in three dimensions} \l{PU_3d}

For the implementation of homogenization formalisms, 
the smooth distribution of 
orientations described in \S\ref{smooth_sec_3d}
is approximated by
 a piecewise-uniform distribution. 
For this purpose, the positive integers $N$ and $M$ are taken to be sufficiently large that the partition \r{A_part_3d} may be
 replaced by the partition
\begin{equation}
 \mathcal{S}_{\mathcal{A}} = 
    \bigcup^{M-1}_{\ell =0}   \bigcup^{N-1}_{k=0}    \mathcal{S}_{\mathcal{A}}^{[\bar{\theta}_k, \bar{\phi}_\ell]},
\end{equation}
where 
$ \mathcal{S}_{\mathcal{A}}^{[\bar{\theta}_k, \bar{\phi}_\ell]}$ represents the subset of  particles of material $\mathcal{A}$  
 with polar orientation angle  $\bar{\theta}_{k} = (\theta_k+\theta_{k+1})/2 $, $k\in\lec0, 1,  \ldots, N-1\ric$, and  azimuthal  orientation angle  $\bar{\phi}_{\ell} = (\phi_\ell+\phi_{\ell+1})/2 $, $\ell\in\lec0, 1,  \ldots, M-1\ric$. 
The volume fraction of 
the particles
 belonging to the subset $\mathcal{S}_{\mathcal{A}}^{\left[ \bar{\theta}_{k}, \bar{\phi}_\ell \right]}$ is denoted by $f_{\mathcal{A}_{k, \ell}}$. 
For this discrete distribution, the probability that a
particle has a polar orientation angle $\bar{\theta}_k$
 and an azimuthal  orientation angle $\bar{\phi}_\ell$
  is given as
\begin{equation}
    \mathrm{P}\left[ \theta = \bar{\theta}_k, \phi = \bar{\phi}_\ell  \right] =
    \frac{  f_{\mathcal{A}_{k, \ell}}}{\fa} =
     g_3 \left( \bar{\theta}_k \right) \, \sin \bar{\theta}_k \, \Delta\theta_k \,
     \Delta\phi_\ell,
\end{equation}
with $\Delta\phi_\ell = \phi_{\ell+1} - \phi_{\ell}$. 
And  $N$ and $M$ are taken to be  sufficiently large that the constraint
\begin{equation}
    \displaystyle  \sum_{\ell=0}^{M-1} \sum_{k=0}^{N-1} g_3 \left( \bar{\theta}_k \right) \,  \sin \bar{\theta}_k \, \Delta\theta_k  \Delta\phi_\ell= 1 
\end{equation}
is satisfied.

\subsection{Sampling of  PDF: 3D orientational distribution}\l{Sample3D}

Next we address the question: At which values  $\bar{\theta}_k \in \le 0, \theta_N  \ri$ and
$\bar{\phi}_{\ell}  \in \le 0, 2 \pi \ri$
 should the PDF $g_3 (\theta,\phi)$ be sampled so that the piecewise-uniform orientational distribution  described in \S\ref{PU_3d}  adequately represents the smooth distribution of
 orientations described in \S\ref{smooth_sec_3d}? 
 
 Since $g_3 (\theta,\phi)\equiv g_3(\theta)$ has been taken to be independent of $\phi$, as specified via Eq.~\r{eq_3Dpdf},
 the  sampling density for $\phi$ should be chosen to be uniform. Then the value of $\Delta\phi_\ell$ is the same for all values of  $\ell \in \lec  0, 1, \ldots M-1 \ric$, regardless of the value of $\phi$ or $\theta$.
 The sampling procedure for $\theta$ follows the inverse  transform sampling method described in \S\ref{Sample2D}, but here based  on the CDF
\begin{eqnarray}
G_3(\theta) &=&\displaystyle\int_{\phi'=0}^{2\pi} \displaystyle\int_{\theta'=0}^{\theta} g_3(\theta') \sin\theta' \mathrm{d}\phi' \mathrm{d}\theta' \nonumber
\\
&=&  \frac{2 \, \mathrm{erf} \left( \frac{i \, \sigma}{\sqrt{2}} \right) - \mathrm{erf} \left( \frac{i \, \sigma^2 + \theta}{\sqrt{2} \, \sigma} \right) - \mathrm{erf} \left( \frac{i \, \sigma^2 - \theta}{\sqrt{2} \, \sigma} \right)}{2 \, \mathrm{erf} \left( \frac{i \, \sigma}{\sqrt{2}} \right) - \mathrm{erf} \left( \frac{i \, \sigma^2 + \theta_N }{\sqrt{2} \, \sigma} \right) - \mathrm{erf} \left( \frac{i \, \sigma^2 - \theta_N}{\sqrt{2} \, \sigma} \right)},
\end{eqnarray}
such that $G_3(0)=0$ and $G_3(\theta_N )=1$.
As in \S\ref{Sample2D}, this procedure results in a
sampling  density that is concentrated most around the maximum of the PDF, 
which is illustrated in Fig.~\ref{Fig3}(b1,b2).

\subsection{Homogenization formalisms: 3D distribution of   orientations} \l{Homogenization3D}

Since the distribution of orientations for the {particles of material $\mathcal{A}$} is taken to be invariant under rotation about the $z$ axis, the
relative permittivity dyadic of the HCM generally has the uniaxial
form
\begin{equation} \l{eps-uniaxial}
\=\eps^{\rm HCM} = \epst^{\rm HCM} \, \le \ux \, \ux +  \uy \, \uy \ri+ \epsz^{\rm HCM} \, \uz \, \uz,
\end{equation}
with
the relative permittivity parameters
 $\epst^{\rm HCM}\in\mathbb{C}$ and $\epsz^{\rm HCM} \in\mathbb{C}$.  The  Bruggeman estimate of $\=\eps^{\rm HCM}$ is written as $\=\eps^{\rm Br}= \epst^{\rm Br} \, \le \ux \, \ux +  \uy \, \uy \ri+ \epsz^{\rm Br} \, \uz \, \uz$ and the Maxwell Garnett estimate of $\=\eps^{\rm HCM}$  
as $\=\eps^{\rm MG}= \epst^{\rm MG} \, \le \ux \, \ux +  \uy \, \uy \ri+ \epsz^{\rm MG} \, \uz \, \uz$. 

\subsubsection{Bruggeman homogenization formalism: 3D orientational distribution}

The  Bruggeman estimate  $\=\eps^{\rm Br}$ is provided implicitly by the nonlinear dyadic equation
\cite{MAEH}
\begin{equation} \label{Br_3d}
\le
\sum^{M-1}_{\ell=0}
\sum^{N-1}_{k=0}  f_{\mathcal{A}_{k,\ell}} \, \=a^{\mathcal{A}_{k,\ell}/ \rm Br} \ri + \fb \, \=a^{\rm \mathcal{B}/Br}=\=0.
\end{equation}
Herein the polarizability density dyadic
\begin{equation}
 \=a^{\mathcal{A}_{k,\ell}/ \rm Br} = \le \eps_\mathcal{A} \=I - \=\eps^{\rm Br} \ri \.   \=P^{\mathcal{A}_{k,\ell}/ \rm Br} , 
 \end{equation}
is defined using the dyadic
 \begin{equation}
 \=P^{\mathcal{A}_{k,\ell}/ \rm Br} =  \les \=I + i \omega \=D^{\mathcal{A}_{k,\ell}/ \rm Br} \. \le \eps_\mathcal{A} \=I - \=\eps^{\rm Br} \ri \ris^{-1} ,
 \end{equation}
which contains the depolarization dyadic $\=D^{\mathcal{A}_{k,\ell}/ \rm Br}$ relevant to
 a particle of material $\mathcal{A}$ with polar orientation angle
$\bar{\theta}_k$ and azimuthal orientation angle $\bar{\phi}_\ell$
 immersed in the  HCM. An expression for this depolarization dyadic is provided in the Appendix.
 The  polarizability density dyadic $\=a^{\rm \mathcal{B}/Br}$ 
 in Eq.~\r{Br_3d}
 is defined in terms of $ \=P^{\rm \mathcal{B}/Br}$ provided in Eqs.~\r{a_2d} and \r{p_2d}.

 A Jacobi iterative scheme~--~which is a generalization of that specified in Eq.~\r{Jacob_2d}~--~is employed to numerically extract  $\=\eps^{\rm Br}$ from Eq.~\r{Br_3d}, as follows.
 The $\le n+1 \ri^{\rm th}$ iterate of  $\=\eps^{\rm Br}$ is delivered in terms of its
  $n^{\rm th}$ iterate as
\begin{eqnarray} \l{Jacob_3d}
    \=\eps^{\mathrm{Br}}[n+1]& =&  \les  \eps_{\mathcal{A}}  \left(\displaystyle
    \sum_{\ell=0}^{M-1}
    \sum_{k=0}^{N-1}  f_{\mathcal{A}_{k, \ell}} \, \={P}^{\mathcal{A}_{k,\ell}/ \rm Br}[n] \right) + \fb \, \eps_{\mathcal{B}} \, \={P}^{\rm \mathcal{B}/Br}[n]  
   \ris 
   \nonumber \\ &&
   \. \les \left( \displaystyle
   \sum_{\ell=0}^{M-1}
   \sum_{k=0}^{N-1} f_{\mathcal{A}_{k, \ell}} \, \={P}^{\mathcal{A}_{k,\ell}/ \rm Br}[n] 
   \right)
   +  \fb \, \={P}^{\rm \mathcal{B}/Br}[n] \ris^{-1},
\end{eqnarray}
wherein
 the dyadics
 $ \={P}^{\mathcal{A}_{k,\ell}/ \rm Br}[n]$ and $ \={P}^{\rm \mathcal{B}/Br}[n]$
 are defined as in Eqs.~\r{PAPB} but with
 the depolarization dyadic
  $ \=D^{{\mathcal{A}_k/ \rm Br}}[n]$ therein replaced by
 $ \=D^{\mathcal{A}_{k,\ell}/ \rm Br}[n]$.
As in \S\ref{Br_sec_2d},
a suitable initial dyadic for the Jacobi scheme is found to be
 $\=\eps^{\mathrm{Br}}[0] = \fa\, \=\eps_{\mathcal{A}} + \fb \,\=\eps_{\mathcal{B}}$. Convergence  for all numerical results presented
 here was typically achieved within 14 iterations.

\subsubsection{Maxwell Garnett homogenization formalism: 3D orientational distribution }

The  Maxwell Garnett formalism estimate of $\=\eps^{\mathrm{HCM}}$ is delivered explicitly by the formula
\cite{MAEH}
\begin{equation}
     \=\eps^{\mathrm{MG}}= \eps_{\mathcal{B}}\, \={I} + \displaystyle
     \sum_{\ell=0}^{M-1}
     \sum_{k=0}^{N-1} \left[ f_{\mathcal{A}_{k, \ell}} \, \={a}^{\mathcal{A}_{k,\ell}/\mathcal{B}} \. \left(\={I} -   \frac{f_{\mathcal{A}_{k, \ell}}}{3 \eps_{\mathcal{B}}} \, \={a}^{\mathcal{A}_{k,\ell}/\mathcal{B}} \right)^{-1} \right],
\end{equation}
wherein 
the polarizability density dyadic
\begin{equation}
  \={a}^{\mathcal{A}_{k, \ell} / \mathcal{B}} = \le \eps_\mathcal{A}- \eps_\mathcal{B} \ri  
   \les \=I + i \omega  \le \eps_\mathcal{A}  - \eps_{\mathcal{B}} \ri \=D^{\mathcal{A}_{k, \ell}/\mathcal{B}}  \ris^{-1}
 \end{equation}
 contains the depolarization dyadic $\=D^{ \mathcal{A}_{k, \ell}/\mathcal{B}}$
relevant to a particle on material $\mathcal{A}$  with polar orientation angle
$\bar{\theta}_k$ and azimuthal orientation angle $\bar{\phi}_\ell$
 immersed in material $\mathcal{B}$. An expression for this depolarization dyadic is provided in the Appendix.

\subsection{Numerical results: 3D distribution of   orientations} \l{Results3D}

In this section, the Bruggeman formalism is applied to estimate the 
constitutive parameters of an HCM arising from a mixture of particles of material $\mathcal{B}$ and
(1) spheroidal, (2) hemispheroidal, or (3) 
doubly-truncated spheroidal {particles of material $\mathcal{A}$. For the Maxwell Garnett formalism,
the shape of the particles of material $\mathcal{B}$ is irrelevant.
The orientations of the  {particles of material $\mathcal{A}$} in three-dimensional space  are characterized by the  Gaussian PDF of Eq.~\r{eq_3Dpdf}.
The relative permittivities chosen for the component materials $\mathcal{A}$ and $\mathcal{B}$ are the same as those chosen in  \S\ref{Results2D}.  The truncation parameter $\kappa=0.1$ for doubly-truncated spheroidal {particles of material $\mathcal{A}$}~--~with  the exception of Fig.~\ref{Fig9} wherein $\kappa$ is varied.
The numbers $N$ and $M$ of sampled values of $\bar{\theta}_k$, {$k\in\lec0, 1, \ldots,N-1\ric$}, and $\bar{\phi}_{\ell}$, {$\ell\in\lec0, 1, \ldots,M-1\ric$},
respectively, were determined by numerical experimentation as described in  \S\ref{Results2D}.
Typically, the tolerances $\delta^{\rm j} < 0.0001$, $\rm{j}\in\lec\rm{Br, \, MG}\ric$,
 were achieved for
 $N< 60$ and $M < 8$ for small values of $\sigma$; for larger values of $\sigma$, smaller values of $N$ and $M$ could be used.

As discussed in \S\ref{Homogenization3D}, the relative permittivity dyadic of the HCM generally takes the uniaxial form given by Eq.~\r{eps-uniaxial}. 
Attention is focussed upon the anisotropy of the  HCM, 
characterized by $\left| \gamma^{\rm HCM} \right| =  \left| \epsz^{\rm HCM} / \epst^{\rm HCM} \right|$, and its relation to: (i) the 
standard deviation $\sigma$ of the orientational distribution of the particles of material $\mathcal{A}$, (ii) the shape of the same
particles, and (iii) the volume fraction $\fa$.
Therefore, the anisotropy factor $\left| \gamma^{\rm j} \right| =   \left| \epsz^{\rm j} / \epst^{\rm j} \right|$,  $\rm{j}\in\lec\rm{Br, \, MG}\ric$,
is plotted against  $\sigma$,
 $\fa$,  $\alpha$, and  $\kappa$, for the  three different shapes of {particles of material $\mathcal{A}$}.
The degree of anisotropy exhibited by the HCM is gauged by the deviation of $ \left|\gamma^{\rm HCM} \right| $ from unity.
Further numerical results, in the form of corresponding plots of the real and imaginary parts of $\epst^{\rm Br}$, $\epsz^{\rm Br}$, $\epst^{\rm MG}$, and $\epsz^{\rm MG}$ as functions of $\sigma$,
 $\fa$,  $\alpha$, and  $\kappa$ are provided in the {\bf Supplementary Material}. 


\begin{figure}[!htb]
\centering
\includegraphics[width=16cm]{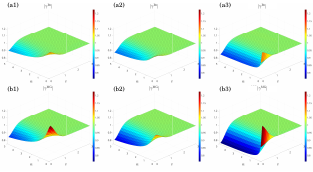} 
  \vspace{0mm}  \hfill
\caption{\label{Fig5} HCM anisotropy factors (a) $ \left| \gamma^{\rm Br} \right| $ and (b) $ \left| \gamma^{\rm MG} \right|  $ versus $\sigma\in[0.1, 3]$ and   $\alpha\in(0,5]$ for (a1, b1) spheroidal, (a2, b2) hemispheroidal, and (a3,b3) doubly-truncated spheroidal ($\kappa=0.1$) {particles of material $\mathcal{A}$}, when $\fa = 0.3$.
}
\end{figure}

The HCM anisotropy factor $ \left| \gamma^{\rm Br} \right| $ is plotted in Figs.~\ref{Fig5}(a1--a3) as a function of  $\sigma \in[0.1, 3]$ and  $\alpha \in (0, 5]$ for the three shapes of {particles of material $\mathcal{A}$}, when $\fa=0.3$.
For $\sigma \gtrsim1$, $ \left| \gamma^{\rm Br} \right| \approx 1 $ regardless of the value of $\alpha$, indicating that the HCM is approximately isotropic for $\sigma \gtrsim 1$. 
The greatest degree of HCM anisotropy occurs in the limit $\sigma \to 0$. 
In this case,  $ \left| \gamma^{\rm Br} \right| $ reaches its maximum as $\alpha \to 0$ and decreases smoothly as $\alpha$ increases.
In the limit $\sigma \to 0$, a substantially higher degree of anisotropy is observed when the {particles of material $\mathcal{A}$} are 
doubly-truncated spheroids with truncation parameter $\kappa=0.1$, as
compared to the cases where the {particles of material $\mathcal{A}$}   are
either spheroids or hemispheroids.
There are only modest differences in $ \left| \gamma^{\rm Br} \right| $ when the 
 particles of material $\mathcal{A}$ are  spheroids and hemispheroids.

The plots of  $ \left| \gamma^{\rm MG} \right|  $  against   $\sigma \in [0.1, 3]$ and $\alpha \in (0, 5]$ in  Figs.~\ref{Fig5}(b1-b3) are qualitatively similar to the corresponding plots of $ \left| \gamma^{\rm Br} \right|$. The Maxwell Garnett formalism predicts rather higher degrees of HCM anisotropy in the regime where $\sigma$ approaches $0$, for both large and small values of $\alpha$.


\begin{figure}[!htb]
\centering
\includegraphics[width=16cm]{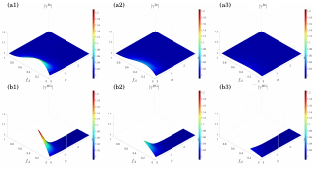} 
  \vspace{0mm}  \hfill
\caption{\label{Fig6} HCM anisotropy factors (a) $ \left| \gamma^{\rm Br} \right|  $ and (b) $ \left| \gamma^{\rm MG} \right|   $ versus   $\sigma\in[0.1, 3]$ and  $\fa\in(0,1)$ for (a1, b1) spheroidal, (a2, b2) hemispheroidal, and (a3,b3) doubly-truncated spheroidal ($\kappa=0.1$) {particles of material $\mathcal{A}$}, when  $\alpha=0.1$.
}
\end{figure}

 In Figs.~\ref{Fig6}(a1--a3),
$\left|\gamma^{\rm Br}\right|$ is plotted against   $\sigma\in[0.1,3]$ and  $\fa\in(0,1)$ for the three shapes of {particles of material $\mathcal{A}$}, when  $\alpha=0.1$.
In the limits $\fa\to 0$ and $\fa \to 1$ we have $\left|\gamma^{\rm Br}\right| \to 1$, regardless of the value of $\sigma$; i.e., the HCM becomes isotropic in the limits $\fa\to 0$ and $\fa \to 1$, as expected \cite{MAEH}.
For $\sigma \gtrsim 1$, $ \left| \gamma^{\rm Br} \right| \approx 1 $ regardless of the value of $\fa$, indicating that the HCM becomes approximately isotropic for $\sigma \gtrsim 1$. 
The greatest degree of HCM anisotropy occurs in the limit $\sigma \to 0$:
 as $\fa$ increases from 0,  $ \left| \gamma^{\rm Br} \right| $ first increases smoothly until it reaches a maximum at $\fa \approx 0.3$, and then it
 decreases smoothly as $\fa$ increases to unity.
In the limit $\sigma \to 0$, a higher degree of HCM anisotropy emerges when the {particles of material $\mathcal{A}$} are   spheroids, as compared to  hemispheroids and doubly-truncated spheroids.
The hemispheroids deliver slightly higher degrees of HCM anisotropy than the 
doubly-truncated spheroids, most conspicuously when $\fa \approx 0.3$ at small values of $\sigma$.

As  in Fig.~\ref{Fig5},
the plots of $ \left| \gamma^{\rm MG} \right| $  against  $\sigma \in [0.1, 3]$ and  $\fa \in (0,0.3]$  in Figs.~\ref{Fig6}(b1--b3) are qualitatively similar to the corresponding plots of $ \left| \gamma^{\rm Br} \right|$ for $\fa \in ( 0,0.3 ]$. In the limit $\sigma \to 0$, the Maxwell Garnett formalism predicts  rather higher degrees of HCM anisotropy as $f_\mathcal{A}$ approaches $0.3$.


\begin{figure}[!htb]
\centering
\includegraphics[width=16cm]{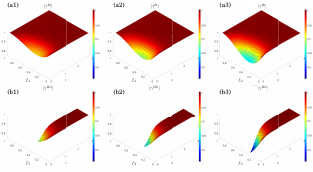} 
  \vspace{0mm}  \hfill
\caption{\label{Fig7}
 HCM anisotropy factors (a) $ \left| \gamma^{\rm Br} \right|  $ and (b) $ \left| \gamma^{\rm MG} \right|   $ versus   $\sigma\in[0.1, 3]$ and  $\fa\in(0,1)$ for (a1, b1) spheroidal, (a2, b2) hemispheroidal, and (a3,b3) doubly-truncated spheroidal ($\kappa=0.1$) {particles of material $\mathcal{A}$}, when  $\alpha=3$.
}
\end{figure}

The plots of Fig.~\ref{Fig6} are repeated in Fig.~\ref{Fig7} but with
  $\alpha=3$.
Similarly to Figs.~\ref{Fig6}(a1--a3), Figs.~\ref{Fig7}(a1--a3) indicate that
the HCM becomes isotropic in the limits $\fa\to 0$ and $\fa \to 1$,
and $ \left| \gamma^{\rm Br} \right| \approx 1 $ regardless of the value of $\fa$   for $\sigma \gtrsim 1$. 
The greatest degree of HCM anisotropy occurs in the limit $\sigma \to 0$; then,
as $\fa$ increases from 0, the anisotropy factor $ \left| \gamma^{\rm Br} \right| $ decreases smoothly until it reaches a minimum at $\fa \approx 0.3$, and thereafter it increases smoothly as $\fa$ increases to unity.
 This is in contrast to the scenario illustrated in Fig.~\ref{Fig6}(a1--a3) wherein
$ \left| \gamma^{\rm Br} \right| $ has a maximum at $\fa \approx 0.3$ for small values of $\sigma$.
In the limit $\sigma \to 0$, slightly higher degrees of anisotropy are predicted when the {particles of material $\mathcal{A}$} are  doubly-truncated spheroids,
as compared to   spheroids and hemispheroids.

As  in Figs.~\ref{Fig5} and \ref{Fig6},
the plots of the HCM anisotropy factor $ \left| \gamma^{\rm MG} \right|  =  \left| \epsz^{\rm MG} / \epst^{\rm MG} \right| $  against the standard deviation $\sigma \in [0.1, 3]$ and volume fraction $\fa \in (0,0.3]$  in Fig.~\ref{Fig7}(b1--b3) are qualitatively similar to the corresponding plots of $ \left| \gamma^{\rm Br} \right|$ for $\fa \in ( 0,0.3 ]$. 
Somewhat higher degrees of HCM anisotropy are predicted by
Maxwell Garnett formalism 
at small values of  $\sigma$ as $f_\mathcal{A}$ approaches $0.3$.


\begin{figure}[!htb]
\centering
\includegraphics[width=16cm]{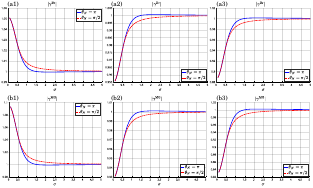} 
  \vspace{0mm}  \hfill
\caption{\label{Fig8} HCM anisotropy factors (a) $ \left| \gamma^{\rm Br} \right|  $ and (b) $ \left| \gamma^{\rm MG} \right|   $ versus  $\sigma\in[0.1, 5]$ for hemispheroids  of material $\mathcal{A}$, when    $\fa=0.3$ and $\theta_N\in\lec\pi/2,\pi\ric$.  Results are presented for  
(a1, b1) $\alpha=0.1$ , (a2, b2) $\alpha=1$, and (a3, b3) $\alpha=3$.}
\end{figure}

Next we turn to the upper bound $\theta_N $  on the  range of 
polar orientation angles $\theta \in \les 0 , \theta_N \ris$ for hemispheroidal {particles of material $\mathcal{A}$}.
As discussed in \S\ref{smooth_sec_3d},
 $\theta_N$ is generally taken to be $\pi$ for these particles as they are not invariant under the mapping $\theta \mapsto \pi - \theta$, 
$\phi \mapsto \phi + \pi$. Therefore, let us determine  the constitutive parameters of the HCM  that would arise if the $\theta$-range 
were $\les 0, \pi/2 \ris$.

The HCM anisotropy factor $\left|\gamma^{\rm Br}\right|$ is plotted against  the standard deviation $\sigma \in [0.1, 5]$ in Figs.~\ref{Fig8}(a1--a3) for hemispheroids of material $\mathcal{A}$}, when   $\fa = 0.3$ and   $\alpha \in \lec 0.1, 1, 3 \ric$.
Results are presented for both  $\theta_N = \pi$ and  $ \theta_N = \pi/2$. 
For the three values of $\alpha$ considered,
$\left|\gamma^{\rm Br}\right|$ converges to unity as $\sigma$ increases beyond 3, for both 
$\theta_N = \pi$ and  $ \theta_N = \pi/2$.
In particular, notice that $\left|\gamma^{\rm Br}\right|$ converging to unity as $\sigma$ increases for $ \theta_N = \pi/2$ indicates that 
the HCM is isotropic when the hemispheroidal particles of material $\mathcal{A}$ are uniformly oriented
 with
$\hat{\#d} \. \uz \geq 0$,
where $\hat{\#d}$ is  indicated in Fig.~\ref{Fig1}(b).

 The greatest degree of HCM anisotropy is achieved as $\sigma \to 0$, for both 
$\theta_N  = \pi$ and  $ \theta_N  = \pi/2$.
For small values of $\alpha$,  $\left|\gamma^{\rm Br}\right| > 1$ as $\sigma \to 0$; in contrast,
 $\left|\gamma^{\rm Br}\right| < 1$ as $\sigma \to 0$
 for large values of $\alpha$.
Clear differences between the degrees of HCM anisotropy for  $\theta_N  = \pi$ and  $ \theta_N = \pi/2$
arise for mid-range values of $\sigma$. For example, at $\sigma = 1.5$ the predicted degree of HCM anisotropy is substantially higher for $\theta_N  = \pi/2$ than for $\theta_N  = \pi$, for all values of $\alpha$ considered.
The degrees of HCM anisotropy
predicted by the Maxwell Garnett formalism in Figs.~\ref{Fig8}(b1--b3) are somewhat
 greater than those predicted by the Bruggeman formalism in Figs.~\ref{Fig8}(a1--a3), most notably for small values of $\sigma$.


\begin{figure}[!htb]
\centering
\includegraphics[width=16cm]{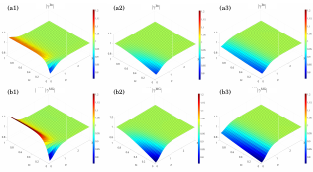} 
  \vspace{0mm}  \hfill
\caption{\label{Fig9} HCM anisotropy factors (a) $ \left| \gamma^{\rm Br} \right|  $ and (b) $ \left| \gamma^{\rm MG} \right|  $ versus   $\sigma\in[0.1, 3]$ and   $\kappa\in(0,1)$ for 
doubly-truncated spheroidal {particles of material $\mathcal{A}$} with equatorial radius (a1, b1) $\alpha=0.1$ , (a2, b2) $\alpha=1$ , and (a3, b3) $\alpha=3$. }
\end{figure}

Lastly, we focus specifically on the HCM arising from doubly-truncated spheroidal {particles of material $\mathcal{A}$}.
The HCM anisotropy factor $\left|\gamma^{\rm Br}\right| $ is plotted against standard deviation $\sigma\in[0.1,3]$ and truncation parameter $\kappa\in(0,1)$ in Figs.~\ref{Fig9}(a1--a3) for doubly-truncated spheroids of material $\mathcal{A}$, when $\fa=0.3$ and   $\alpha \in \lec 0.1, 1, 3 \ric$.
When   $\sigma \gtrsim 1$,  $ \left| \gamma^{\rm Br} \right| \approx 1 $ for all values of $\kappa$ and 
$\alpha$.
The greatest degree of HCM anisotropy occurs in the limit $\sigma \to 0$, when $ \left| \gamma^{\rm Br} \right| $ increases smoothly
  as $\kappa$ increases from 0. 
When $\alpha = 0.1$,  $ \left| \gamma^{\rm Br} \right| $ transitions from being less than unity to greater than unity as $\kappa$ increases from $0$ to $1$.
When $\alpha = 1$, we recover the case of  doubly-truncated spherical
particles of material $\mathcal{A}$; in this case,  the  particle shape approaches that of a sphere as the truncation parameter $\kappa \to 1$, and  $ \left| \gamma^{\rm Br} \right| \to 1$. This indicates that the HCM becomes isotropic in the limit $\kappa \to 1$ for $\alpha = 1$, regardless of the value of $\sigma$, as expected \cite{MAEH}.

The plots of  $ \left| \gamma^{\rm MG} \right|   $  versus $\sigma\in[0.1,3]$ and   $\kappa\in(0,1)$ in Figs.~\ref{Fig9}(b1--b3) are qualitatively similar to the corresponding plots of $ \left| \gamma^{\rm Br} \right|$  in Figs.~\ref{Fig9}(a1--a3). In the limit $\sigma \to 0$, the Maxwell Garnett formalism predicts a rather higher degree of HCM anisotropy than the Bruggeman formalism does.

\section{Discussion and Conclusion}

Advancements in composite-material technology have heightened the need for more sophisticated homogenization formalisms to determine the effective constitutive parameters of  particulate composite materials with complex micro-morphology.
In particular, homogenization formalisms capable of accommodating component particles with complex shapes 
 and non-uniform distributions are required. 
Recent progress has resulted in 
implementations of the Maxwell Garnett formalism
\cite{ML_WRCM} and Bruggeman formalism \cite{TruncatedSpheroids}
for HCMs based on component particles shaped as
 hemispheroids and doubly-truncated spheroids. As is often the case with  homogenization research, these component particles were assumed to be randomly distributed but all aligned in the same direction.
In the preceding sections implementations of the Bruggeman and  Maxwell Garnett formalisms were
developed for
 component particles with
 orientations specified by a  PDF. The
 component particles considered had
  spheroidal, hemispheroidal, and doubly-truncated spheroidal shapes.
  While the PDF selected here was a truncated Gaussian PDF, the 
  piecewise-uniform approach adopted 
  could be readily adapted to a wide range of PDFs.

  By varying the standard deviation $\sigma$ of the orientation distribution of the component particles, the degree of anisotropy exhibited by the HCM can be controlled. Numerical studies based on  spheroidal
  component
   particles 
   made from realistic isotropic dielectric materials,
   with orientations distributed in   a plane,
   revealed that the corresponding HCM is in general a  biaxial dielectric material, but it becomes a uniaxial dielectric material  in the limits  $\sigma \to 0$ and $\sigma \to \infty$. 
   Further
   numerical studies based on  spheroidal, hemispheroidal, and doubly-truncated spheroidal
  component
   particles, 
   with  particle orientations distributed in  three dimensions,
   revealed that the 
   corresponding
   HCM is in general a  uniaxial dielectric material, but the HCM becomes an isotropic dielectric material  in the limit $\sigma \to \infty$; and the degree of anisotropy exhibited by the HCM increases steadily as $\sigma$ decreases.
   
   The values of the constitutive parameters and
   anisotropy factors
    of the HCM
 estimated by the Bruggeman and Maxwell Garnett formalisms 
 turned out to be in broad agreement, 
 over much of the volume-fraction range appropriate to the Maxwell Garnett formalism,
 with the
  Maxwell Garnett estimates of the HCM anisotropy being slightly greater   than the Bruggeman estimates 
  at small values of $\sigma$, 
  while the volume fraction $\fa$ approaches 0.3.

Lastly, 
following the usual practice for homogenization formalisms,
 all the component {particles of material $\mathcal{A}$} considered here share the same shape, 
 as do the all the component  particles of material $\mathcal{B}$. 
  The piecewise-uniform approach adopted 
  could be further adapted to take into account 
   distributions of shapes for the  component particles of either material or both materials.

\section*{Appendix~--~Depolarization dyadics}

Depolarization dyadics~--~which represent integrated singularities of 
dyadic Green functions corresponding to the shapes of  convex particles immersed in homogeneous ambient mediums~--~are key mathematical entities in the Bruggeman and Maxwell Garnett homogenization formalisms (as well as many other less well-known homogenization formalisms). For the implementations of the Bruggeman and Maxwell Garnett formalisms presented in this paper, depolarization dyadics are required relevant to
particles shaped as spheres,
spheroids, hemispheroids, and doubly-truncated spheroids, immersed in isotropic, uniaxial and biaxial dielectric ambient mediums. 
 Each  shape is based on a spheroid with unit polar radius  and equatorial radius $\alpha$, with rotational symmetry axis oriented 
 parallel to the unit vector $\#{\hat{d}}$, as shown in Fig.~\ref{Fig1}. 
While expressions for these depolarization dyadics are available in various scientific publications, for completeness and  convenience these expressions are compiled in this Appendix. 

The depolarization dyadics are denoted by $\=D$ and the relative permittivity dyadic of the homogeneous ambient material by $\=\eps$; in the case  of  the Bruggeman formalism
$\=\eps =  \=\eps^{\rm Br} $ and in the case of the Maxwell Garnett formalism $\=\eps = 
\eps_\mathcal{B} \=I $.


The most general case considered here for the two-dimensional orientational distribution of particles of material 
$\mathcal{A}$   involves a spheroidal particle immersed in a biaxial  ambient medium.
 If the spheroidal particle is centred at the origin of the coordinate system, then its
  surface  is prescribed by the position vector $  \={U} \. \#{\hat{r}}$, where
$ \#{\hat{r}}$ is the unit vector on the surface of the unit sphere and the shape dyadic
\begin{equation}\label{eq_ShapeU}
     \={U} = \alpha \,\={I} + \left( 1 - \alpha \right) \#{\hat{d}} \, \#{\hat{d}}.
\end{equation}
Notice that $\=U = \=I$ in the degenerate case of a spherical particle,
The  corresponding depolarization dyadic is delivered in terms of the double integral  \cite{MW97}
\begin{equation}\label{eq_DepolarizationD}
\=D   =\frac{1}{4 \pi i \omega} \={U}^{-1} \. \left(\int_{\phi_q=0}^{2\pi} 
 \int_{\theta_q = 0}^{\pi} 
\frac{\hat{\#{q}} \,  \hat{\#{q}} \, \sin\theta_q }{\hat{\#{q}} \.  \={U}^{-1}  \. \=\eps \. \={U}^{-1}  \. \hat{\#{q}} }
\mathrm{d}\theta_q \,  \mathrm{d}\phi_q \right) \.  \={U}^{-1} ,
\end{equation}   
with the unit vector $\hat{\#{q}}=  \ux \sin\theta_q \, \cos\phi_q+\uy \sin\theta_q \, \sin\phi_q+\uz \cos\theta_q$. 
The double integration on the right side of Eq.~\r{eq_DepolarizationD} can be expressed in terms of incomplete elliptic integrals of the first and second kinds \cite{W98}; also, it is amenable to evaluation by standard numerical methods.


The most general case considered here for the three-dimensional orientational distribution of particles of material $\mathcal{A}$   involves a convex particle immersed in a uniaxial  ambient medium
characterized by the relative permittivity dyadic
$\=\eps= \epst \, \le \ux \, \ux +  \uy \, \uy \ri+ \epsz \, \uz \, \uz$.
The corresponding depolarization dyadic
\begin{equation}
\=D = \frac{\gamma}{i \, \omega } \=L \. \={\eps}^{-1},
\end{equation}
with $\gamma=\epsz / \epst$,
is related to the dyadic
\begin{equation} \l{Q2_uniaxial}
\=L = \Lt \left(  \=I - \#{\hat{d}} \, \#{\hat{d}}\right)   + \le \frac{1}{\gamma} - 2 \, \Lt \ri \#{\hat{d}} \, \#{\hat{d}}.
\end{equation}
Closed-form expressions for the scalar $\Lt$ are available for spheroidal, hemispheroidal, and doubly-truncated spheroidal shapes as follows:
\begin{itemize}
\item[1.] For the spheroidal  shape \cite{M97},
\begin{equation}\label{eq_LtSspheroid}
\Lt = \frac{1}{2} \left( \frac{1}{\gamma-\alpha^2 \, \gamma} + \frac{\alpha^2 \, \mathrm{sec}^{-1}(\alpha \, \sqrt{\gamma})}{(\alpha^2 \, \gamma-1)^{3/2}} \right).
\end{equation}
\item[2.]
For the hemispheroidal  shape \cite{ML_TAP},
 \begin{multline}\label{eq_Lt1Hemispheroid} 
\Lt =\frac{1}{4 \, h^3}\left\{ \frac{h \, \left( f(1+\alpha^4) + g \, p (1+\alpha \, f) - \alpha^2 (f + 2 \, \gamma \, f + \gamma \, g \, p)\right)}{\gamma \, g \left( 1 - (1 + \gamma)\alpha^2 + \alpha^4 \right)} \right.\\ - \left. \alpha^2 \left( \mathrm{coth}^{-1} \left( \frac{g \, h}{f} \right) - \mathrm{tanh}^{-1} \left( \frac{\alpha(\gamma \, \alpha + f) - 1}{p \, h} \right) \right) \right\} 
 \end{multline}
when $\alpha \leq 1/\sqrt{2} $ with  $f = \sqrt{1-\alpha^2}$, $g=\sqrt{1+\gamma-\alpha^2}$, $h = \sqrt{1 - \gamma \, \alpha^2}$ and $p=\sqrt{1+\alpha^2-\alpha^4-2 \, \alpha \, f}$;  and 
\begin{equation}\label{eq_Lt2Hemispheroid}
\Lt =\frac{v \, (3 - 2 \alpha^2 \, \gamma + 5 - 2 \, v) + 5}{4 \, v \gamma \left[ 3 + \alpha^2 \, \gamma (4 \, \alpha^2 \, \gamma - 7) \right]} + \frac{\alpha^2}{2 \, w^3}\left[ \tan^{-1} \left( \frac{v - 1}{2  \, w} \right) - \mathrm{cot}^{-1}(w \, v) \right]
\end{equation}
when $\alpha>1/\sqrt{2}$ with $v=\sqrt{1+4 \, \alpha^2 \, \gamma}$ and $w=\sqrt{\alpha^2 \, \gamma - 1}$.
\item[3.] For the doubly-truncated spheroidal shape \cite{ML_TAP},
\begin{equation}  \label{eq_LtDTSpheroid}
\Lt = \frac{1}{2 \, (\alpha^2 \gamma - 1)^{3/2}}\left[ \alpha^2 \, \tan^{-1} \left( \kappa \sqrt{\frac{\alpha^2 \, \gamma - 1}{\kappa^2 + \alpha^2 \, \gamma - \alpha^2 \, \kappa^2 \, \gamma}} \right) - \frac{\kappa}{\gamma} \sqrt{\frac{\alpha^2 \, \gamma - 1}{\kappa^2 + \alpha^2 \, \gamma - \alpha^2 \, \kappa^2 \, \gamma}} \, \right].
 \end{equation}
 when  the symmetrically positioned
 truncation planes
 are separated by $2\kappa$, per Fig.~\ref{Fig1}(c). 
 
\end{itemize}
\vspace{10mm}
 
\section*{Acknowledgments}
TGM was partially supported  by
EPSRC (grant number EP/V046322/1).  AL's research   was partially supported by an Evan Pugh University Professorship  at Penn State.

\section*{Declaration of competing interest}
The authors declare that they have no known competing financial interests or personal relationships that could have
appeared to influence the work reported in this paper.

\section*{Data availability}
All data  analyzed in this paper are included in this paper and supplementary material.

\end{document}